# Thermophotovoltaic conversion efficiency measurement at high view factors

Esther López[*,1], Irene Artacho[1] and Alejandro Datas[1,**]

[1]Instituto de Energía Solar – Universidad Politécnica de Madrid, Avenida Complutense 30, 28040 Madrid, Spain

[*]Corresponding author 1: esther.lopez.estrada@upm.es
[**] Corresponding author 2: a.datas@upm.es

**Abstract**

A standardized method for measuring thermophotovoltaic (TPV) efficiency has not been yet established, which makes the reported results difficult to compare. Besides, most of the TPV efficiencies reported to date have been obtained using small view factors, i.e., large cell-to-emitter distances, so the impact of the series resistance is usually underestimated, and the optical cavity effects, i.e., the multiple reflections taking place between the emitter and the cell, are not accounted for experimentally. In this work, we present an experimental setup able to measure the TPV efficiency under high view factors (up to 0.98), by using small emitter-to-cell distances (< 1 mm). This allows a more accurate direct measurement of the TPV efficiency at higher power densities than previous works. As a result, a TPV efficiency of 26.4±0.1 % and a power density of 4.3±0.8 W/cm$^2$ have been obtained for an InGaAs TPV cell with a back surface reflector irradiated by a graphite thermal emitter at 1592 °C.

## 1. INTRODUCTION

Thermophotovoltaics (TPV) is the direct conversion of radiant heat into electricity through the photovoltaic effect. A TPV device includes an incandescent thermal emitter that radiates photons towards a closely spaced TPV cell, which subsequently produces electricity. Typically, a TPV cell is made of a low bandgap semiconductor (e.g., InGaAs at 0.74 eV) able to produce electron-hole pairs from the absorption of low energetic photons whose energy is larger than that of the semiconductor bandgap. By using spectral control elements (e.g., filters or mirrors), sub-bandgap energy photons are reflected and turned back to the emitter to be reabsorbed, and thus, do not account for the energy losses. In this way, the thermophotovoltaic (TPV) conversion efficiency is defined as [1]:

$$\eta_{TPV} = \frac{P_{ele}}{P_{IN} - P_R} = \frac{P_{ele}}{P_{abs}} \quad (1)$$

where $P_{ele}$ is the electrical power generated by the cell; $P_{IN}$ is the incident power on the cell and $P_R$ is the reflected power by the cell, so $P_{abs}$ is the net power absorbed in the cell as a result. In equation (1), $P_R$ is subtracted because it is not considered an energy loss. However, this is only true when the view factor (VF) between the emitter and the cell is close to unity, so the power reflected by the cell can be intercepted by the emitter to be recycled. If the emitter is not a perfect black body (its emissivity $\varepsilon_e$ is lower than 1), then multiple reflections between the emitter and the cell can take place, resulting in a net radiative heat flux from the emitter to the cells [2]. Accounting for the energy losses related to the parasitic absorption resulted from those multiple reflections is essential to determine the value of the denominator in equation (1). Experimentally, this is only possible whether high VFs between the emitter and the cell are established when measuring the TPV efficiency.

In a practical TPV system, the emitter and the cells are integrated in an optical cavity, and the complex radiative exchange between them determines the net absorbed power in the cells that should be used in equation (1). However, the experimental setups developed so far to measure the TPV conversion efficiency use simplified arrangements where the TPV cell is located relatively far away from a thermal emitter with a low view factor. This precludes the establishment of the multiple reflections that take place in a practical TPV system, and thus, avoid the obtention of an accurate efficiency measurement. To circumvent this problem, previous studies either assume a black body emitter (i.e. multiple reflections do not take place) or estimate the net radiative heat flux resulted from the multiple reflections using a theoretical effective emissivity [3], [4]. The latter approach implies assuming an ideal optical cavity without spillage losses, and an apparent view factor < 1 (cells area larger than emitter area) that must be also estimated, increasing the inaccuracy of the results. Moreover, calculating that effective emissivity requires a priori knowledge of the optical properties of the involved surfaces (the cell and the emitter). Different research groups have used different methods to calculate these parameters [3], [5], [6], which make the reported results more difficult to compare. Besides, these indirect measurements are conducted under conditions (temperature, angle of incidence, and spectral range) which may differ significantly with respect to the real operating conditions. Therefore, the accuracy associated to TPV efficiencies obtained with large emitter-to-cell distances (low view factors) might be difficult to determine, which may give misleading results.

Using low VFs has an additional disadvantage: it leads to low power densities, which is the figure of merit of several TPV applications [7], [8]. Besides, measuring the TPV conversion efficiency when generating low power densities may underestimate series resistance issues, which can appear once high VFs are stablished in the final TPV systems and higher power densities are generated as consequence. This can be solved by increasing the size of the emitter to increase the apparent VF of the cell during the TPV conversion efficiency measurement. However, if the emitter and the cell are place at a high distance, multiple reflections are neither considered experimentally. Only by using black body emitters (with $\varepsilon_e \sim 1$) these effects may be neglected in this case.

In order to obtain the TPV conversion efficiency without the need of assuming blackbody emitters nor calculating an effective emissivity from indirect methods, it is necessary to assure that multiple reflections effectively take place between the emitter and the cell. This is possible if we use small emitter-to-cell distances, so the VF can be close to unity. In this work we show an experimental setup where small emitter-to-cell distances between 0.9-0.4 mm are established, resulting in VFs in the range of 0.91-0.98. Importantly, we show that the use of high VFs results in much higher power densities when compared to previous results reported in the literature. The aim of this study is to demonstrate an experimental method for the direct measurement of TPV efficiency that can be applied to any cell/emitter configuration without the need of a priori knowledge or assumption of their optical properties.

1. **METHODS**

In order to obtain the net power absorbed by the cell used in equation (1), an energy balance must be established, according to which the power absorbed by the cell is converted to either electrical power or dissipated heat ($Q$), so the TPV conversion efficiency can be expressed as:

$$\eta_{TPV} = \frac{P_{ele}}{P_{abs}} = \frac{P_{ele}}{P_{ele}+Q} \quad (2)$$

As a consequence, the TPV conversion efficiency is calculated from equation (2), using only two in-situ measurable parameters: $P_{ele}$ (using the 4-wire method with a Keithley 2602B) and $Q$ (using a calorimeter, see APENDIX). Equation (2) includes the complex radiative exchange between the emitter and the cells because a configuration with a high VFs (close to unity) is used. In particular, VFs in the range of 0.91-0.98 are attained because cells with designated area of 0.7 cm$^2$ are located at a distance of 0.9-0.4 mm from an emitter surface of 0.4 cm$^2$. These VF values are calculated from the geometry of unequal square plates [9]. Such small gap between the emitter and the cell can be successfully obtained because, as it is illustrated in Figure 1a, the same structure that holds the emitter over the cell also includes the probes that contact the device. Any other form of front-contact configuration would impede such high VFs. Note that this setup configuration requires fabricating a holder structure and an emitter with specific dimensions adapted to the size of the TPV cell under study.

As counterpart, the mentioned holder must be designed to avoid overheating of the contact probe, so the emitter needs to be thermally isolated inside (see Figure 1a). To this end, the holder is electroplated with gold, so the radiation emitted laterally by the emitter can be reflected. The heat transfer by conduction is also reduced by securing the emitter with ceramic thermal isolators. Besides, the emitter is placed inside a vacuum chamber, which suppresses heat transfer by convection. This way, the emitter can be heated to high temperatures while keeping the contact probes at room temperature, by means of water coolers that refrigerate the surroundings of the holder.

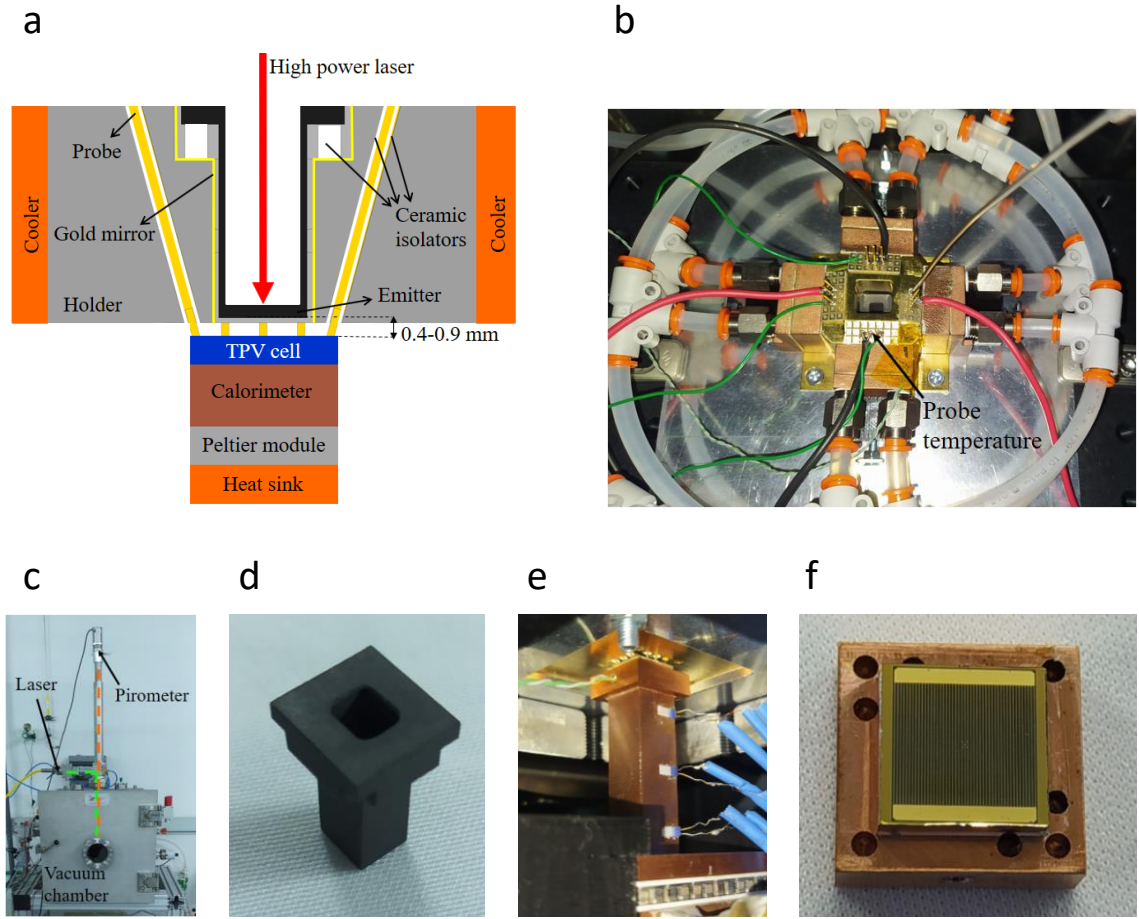

Figure 1: (a) Sketch of the main elements of the experimental setup. (b) Top view of the setup, showing the refrigeration system to control the temperature of the contact probes. (c) Vacuum chamber of the experimental setup and illustration of the optical path follows by the laser beam and the radiation detected by the pyrometer. (d) Graphite emitter with an opened cavity, where the laser spot and the pyrometer spot impinge. (e) Lateral view of the calorimeter, showing the three PT100 temperature sensors (f) InGaAs cell with gold reflector provided by Antora Energy mounted on top of the calorimeter.

The thermal isolation of the emitter, avoiding any physical contact with it, contributes to homogenize its temperature and therefore to increase the accuracy of this challenging measurement. For this reason, the emitter is heated up by wireless power transmission using the light of a RFLA500D laser (500 W, 915 nm, continuous-wave), and the emitter temperature is measured with an optical pyrometer E2MH-F2 (spectral range of 1.6 µm). Both the laser and the pyrometer spots impinge inside a hole opened in the graphite emitter piece (see Figure 1d). This hole acts as an optical cavity, facilitating the temperature homogenization and allowing us to measure the temperature just 2 mm away from the emitter surface that irradiates the TPV cell. Note that the accuracy in the emitter temperature is necessary to provide a fair comparison with other reported results, but this parameter is not used to obtain the TPV conversion efficiency in this work.

The temperature of the TPV cell and the contact probes is controlled during the measurements to be equal to the room temperature. This is important to guarantee that there is no heat leakage through the contact probes, and all the heat generated in the cell flows down through a copper structure where it is quantified (calorimeter). For the cell temperature, we use a PID controller (model P116) connected to a Peltier cooler (model ET-071-20-15-RS) located below the calorimeter, so a heat flux through this structure is forced to reach the room temperature. To control the temperature of the contact probes, the perimeter of the holder is refrigerated with coolers connected to a water chiller (model HRS060-AF-20), whose temperature can be adjusted until the temperature of the probes is equal to the room temperature. Using k-type thermocouples, the temperature of the cell is measured just below it, while the temperature of the probes is measured in one side of the holder (see Figure 1b). This last temperature measurement is representative for all the probes because all of them are homogenously refrigerated: the four sides of the holder are refrigerated with four identical coolers that are connected in parallel, so the heat transfer through every side is expected to be the same. Notice that the probes are in direct physical contact with the holder structure only at the point where its temperature is measured. In any other location, the probes are surrounded by alumina tubes with an inner diameter that is larger than that of the probes. Moreover, the outer diameter of the alumina tubes is also slightly smaller than the diameter of the holes of the holder where the probes are inserted. Thus, this arrangement avoids a thermal contact between the probes and the holding structure. Even if an unintentional punctual contact between the probes and the alumina tubes happened along the holes, the low thermal conductivity of alumina (25 $Wm^{-1}K^{-1}$) together with a high thermal contact resistance (between the holder and the alumina tubes and between the alumina tubes and the probes) would impede significant additional heat fluxes to take place.

To obtain the TPV conversion efficiency, the electrical power generated by the cell and the heat dissipated from the cell will be measured under steady-state conditions, since only then there is an energy balance and equation (2) is fulfilled. Because biasing the cell at different voltages to obtain complete current-voltage characteristics affects the energy balance, during the TPV conversion efficiency measurement the cell will be constantly biased at the maximum power point [10].

## 2. RESULTS AND DISCUSSION

InGaAs TPV cells with back gold reflectors, provided by Antora Energy, have been tested in the experimental setup described above (see Figure 1). The results of the measurements carried out under different illumination conditions, corresponding to different emitter temperatures, are summarized in Table I and II. The uncertainties that appear in these tables correspond to the measurement variation ranges. For some parameters, like the voltage and the power at the maximum power point, the uncertainty can be neglected ($\Delta V_{mpp} < 8 \cdot 10^{-4}$ V and $\Delta P_{mpp} < 6 \cdot 10^{-3}$ W). However, special attention is required to analyze the uncertainty of the electrical power density generated by the cell. Although this is usually dominated by the uncertainty in the designated area of the devices, in the experimental setup used this work there is a more important source of error. This is because the holder structure that contains the emitter and the contact probes (Figure 1a) acts as a partial mask. A complete mask is only possible when the mask element completely touches the devices, especially when high VF are used. Because this is not the case in our setup, the value of the effective illuminated area is somewhere in between the area of the aperture that contains the emitter (0.5 cm$^2$) and the designated area of the cell (0.7 cm$^2$). As a result, the value of the electrical power density (PD) is delimited between a maximum and a minimum limit, respectively.

Table I: List of the parameters measured to obtain the electrical power density (PD) and the TPV efficiency ($\eta_{TPV}$) of the InGaAs cells tested in this work.

| $T_{emitter}$ (ºC) | Impp (A) | Vmpp (V) | Pmpp (W) | $PD_{min}$ (W/cm$^2$) | $PD_{Max}$ (W/cm$^2$) | Q (W) | $\eta_{TPV}$ (%) |
|---|---|---|---|---|---|---|---|
| 1154±2 | 1.109±0.001 | 0.396 | 0.439 | 0.6 | 0.9 | 1.45±0.07 | 23.2±0.9 |
| 1294±3 | 2.128±0.002 | 0.396 | 0.842 | 1.2 | 1.7 | 2.55±0.06 | 24.8±0.4 |
| 1427±5 | 3.48±0.01 | 0.422 | 1.469 | 2.1 | 3.0 | 4.06±0.05 | 26.6±0.2 |
| 1480±3 | 4.114±0.005 | 0.431 | 1.771 | 2.5 | 3.6 | 4.83±0.01 | 26.8±0.1 |
| 1562±4 | 5.283±0.002 | 0.438 | 2.316 | 3.3 | 4.7 | 6.35±0.02 | 26.7±0.1 |
| 1592±6 | 6.04±0.02 | 0.416 | 2.515 | 3.5 | 5.1 | 7.00±0.04 | 26.4±0.1 |

Table II: Average value of the temperature measurements obtained under steady-state conditions.

| $T_{emitter}$ (ºC) | $T_{room}$ (ºC) | $T_{probes}$ (ºC) | $T_{cell}$ (ºC) |
|---|---|---|---|
| 1154±2 | 30.3 | 30 | 29.9 |
| 1294±3 | 30.4 | 30 | 29.9 |
| 1427±5 | 30.5 | 30 | 29.9 |
| 1480±3 | 30.6 | 31 | 30.5 |
| 1562±4 | 30.7 | 31 | 31.0 |
| 1592±6 | 30.8 | 31 | 30.9 |

The TPV efficiencies and electrical power densities obtained in this work are plotted in Figure 2, together with the record efficiencies reported in the literature [3], [5], [6], [11].

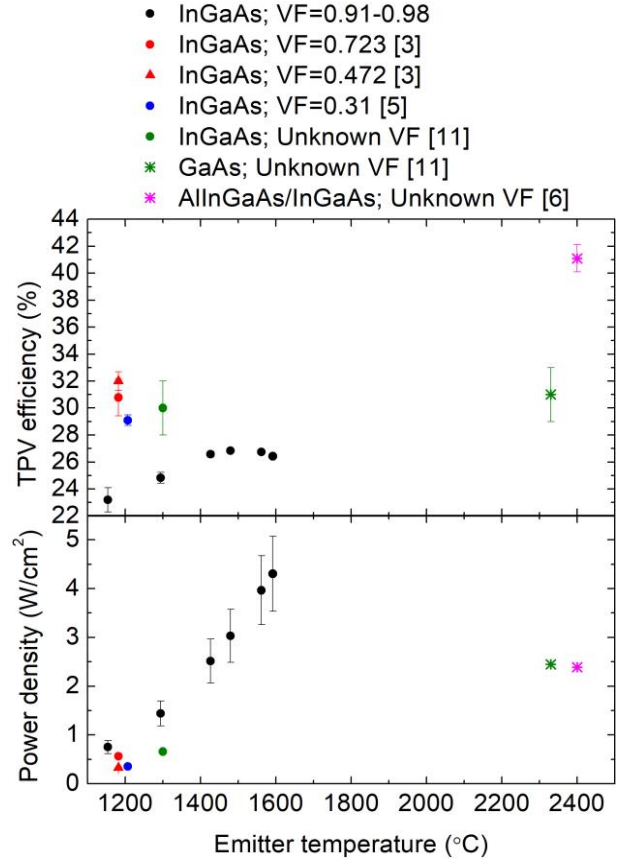

Figure 2: (top) TPV efficiency and (bottom) electrical power density generated by different TPV cells, including the InGaAs cells with back reflectors tested in this work (black dots). The VF used in every case is specified in the legend [3], [5], [6], [11].

The TPV efficiencies obtained in this work (black dots in Figure 2) show a maximum of 26.8 % for an emitter temperature of 1480 ºC. At lower emitter temperatures the TPV efficiency is reduced due to a higher misalignment between the spectral response of the cell and the spectrum illumination, while at higher temperatures it is reduced due to series resistance effects. This last effect is only visible when high current densities (or analogously, high power densities) are generated, so depending on whether high or low VFs are used, this effect will be visualized or hidden respectively. Therefore, just by reducing the VF the TPV efficiency can be increased, since it is possible to benefit from a better spectral alignment at higher emitter temperatures without the corresponding electrical losses. This can be noticed in the results of Ref. [3], where a higher TPV efficiency of 32 % is obtained when operating with a smaller VF of 0.472 (red triangle in Figure 2), compared to a lower TPV efficiency of 30.78 % obtained with a higher VF of 0.723 (red dot in Figure 2).

However, for most of the final TPV applications high VFs, close to unity, are required, so it is important to measure the TPV efficiency under these conditions. This is the case of the TPV efficiencies reported in this work, which are obtained under high VF of 0.91-0.98 by using small emitter-to-cell distance. This configuration ensures that multiple reflections can take place between the emitter and the cell, so the TPV efficiency can be obtained without requiring numerical extrapolations and accounting for the associated energy losses. Besides, this allows us to measure the TPV efficiency while producing higher power densities compared to previous works. For example, a TPV efficiency of 26.4 % is obtained when producing more than 4.3 W/cm$^2$ for an emitter temperature of 1592 ºC, which is a remarkable result. As far as we know, only works focused on measuring the efficiency of TPV systems [12] (which is not the as the TPV efficiency) have been able to produce power densities comparable to the ones reported in this work for equivalent emitter temperatures. It is worth noticing that the cells tested in this work, as the ones used in most previous results, lack of antireflective coating. So, incorporating antireflective coatings would increase even more the generated power density. A higher power density would be also obtained by increasing the temperature of the emitter (above 1592 ºC), however the produced current would exceed the current limit (6 A) of the source-meter that is used (Keithley 2602B). Finally, it is also important to note that because in this work the cell area is partially masked the effects of series resistance are slightly softener respect to equivalent cases where the full area would be illuminated.

3. **CONCLUSIONS**

In this work, an experimental setup has been implemented to measure TPV efficiencies under high view factor (0.91-0.98). This is important to obtain TPV conversion efficiency through direct measurements, without the need of using perfect black body emitters or calculating an effective emissivity from indirect methods. In the setup presented in this work, the formation of an optical cavity between the emitter and the cell by using small emitter-to-cell distance, allows to provide a more direct TPV efficiency measurement, reaching a maximum of 26.8 % for an emitter temperature of 1480 ºC. Besides, the use of high VFs allows the production of higher power densities compared to previous works, reaching more than 4.3 W/cm$^2$ for an emitter temperature of 1592 ºC, which is a remarkable result.

## A. APPENDIX

### A.1 Calibration of the heat flux measurement

In a solid bar like the structure used as calorimeter (Figure 1e), the heat that flows through this element is proportional to the temperature gradient established in it, following the equation (3):

$$Q = k \cdot \Delta T \quad (3)$$

where $k$ is a constant that depends on the features of the solid bar. In order to avoid systematic errors, the relation between the heat that flows through the calorimeter and its temperature gradient has been calibrated. To this end, we have used two different devices: a planar resistor and a completely metallized GaSb cell that cannot emit luminescence photons. Each of these devices has been located on top of the calorimeter structure, dissipating a known electrical power as heat. The temperature of these devices has been controlled to be equal to the room temperature. Then, under vacuum and steady-state conditions, the resulted energy balance implies that the heat that flows through the calorimeter is equal to the electrical power dissipated in these devices. The temperature gradient resulted in the calorimeter is measured with PT100 sensors in 3-wire connection, using a NI 9217 to minimize the measurement uncertainty. These sensors are located at three equidistant positions (the T1 sensor on the top, the T2 sensor in the middle and the T3 sensor on the bottom). Figure 3 shows the relation between the dissipated electrical power and the temperature gradient evaluated at different positions under steady-state conditions.

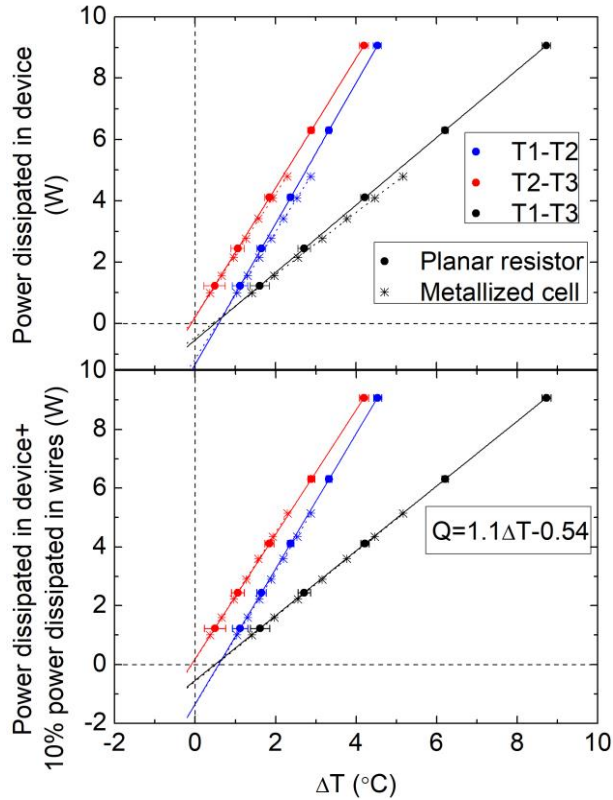

Figure 3: (top) Linear relation between the temperature gradient at different positions of the calorimeter and the power dissipated in a planar resistor and a metalized cell. (bottom) Adding 10% of the power

dissipated in the wires to the power dissipated in the device terminals, similar curves are obtained for the two devices. From the black trend line, the relation between the heat flux and the temperature gradient in the calorimeter is obtained.

The electrical power dissipated in the terminals of the two devices used to calibrate the calorimeter is measured using the 4-wire method with a Keithley 2602B. While in the planar resistor this power is characterized by high voltages and low currents, in the completely metallized GaSb cell it is characterized by high currents and low voltages. As a result, during the measurements with the metallized GaSb cell, a higher power is dissipated in the wires that connect the device with the walls of the vacuum chamber, which are also at room temperature. A portion of this power can contribute to the heat measurement, since part of the heat generated in the wires can be transfer to the device and therefore to the calorimeter, producing a higher temperature gradient in it. This is observed in Figure 3 (top). For this reason, the electrical power dissipated in the wires has been also measured with the 4-wire method. In order to determine what is the percentage of this power that is transferred to the calorimeter, we have assumed that this percentage remains constant, regardless of whether we use the planar resistor or the metallized GaSb cell, since the wire connections do not change from case to case. Then, we have added different percentages of the power dissipated in the wires until both devices provide the same trend lines. This is illustrated in Figure 3 (bottom), where adding a 10% of the power dissipated in the wires is necessary to obtain the same trend lines from the two devices used for the calibration. This suggests that the relation between the heat that flows through the calorimeter and its temperature gradient is the one expressed by the trend lines of this figure. Although, according to equation (3), these trend lines should have a zero interception, systematic measurement error in the PT100 sensors leads to deviations from the theoretical dependency. The bigger separation between the sensors, the lower relative error in the temperature gradient. This is why in this work the value of the heat flux is obtained from the black trend line in Figure 3 (bottom). The uncertainty associated to this trend line is lower than the variation range of the heat measurements, this is why this variation range is represented as uncertainty in Table I.

**A.2 Calibration of the emitter temperature measurement**

The temperature measurement obtained with the pyrometer (E2MH-F2) depends not only on the emissivity of the material used to fabricate the emitter, but also on the geometry of the emitter's cavity where the spot of the pyrometer impinges (Figure 1d). Besides, because before to reach the pyrometer the radiation passes through an anti-reflective window placed in the top of the vacuum chamber and a dichroic reflector (see Figure 1c), the temperature measurement also depends on the transmittance of these interfaces. All these effects can be taken into account by setting both, the emissivity and transmittance parameters of the pyrometer, or just one of them, since the effect of both parameters in the temperature measurement is the same.

In order to determine the value of the parameters that should be set in the pyrometer to obtain correct temperature measurements, a calibration piece was fabricated using the same material as the emitter (isostatic graphite). This calibration piece had the same upper cavity as the emitter, but it was bigger, so there was room to introduce a thermocouple into a hole placed 2 mm below the bottom of the cavity, at the same distance as the surface of the emitter that faces the TPV cell (see Figure 4).

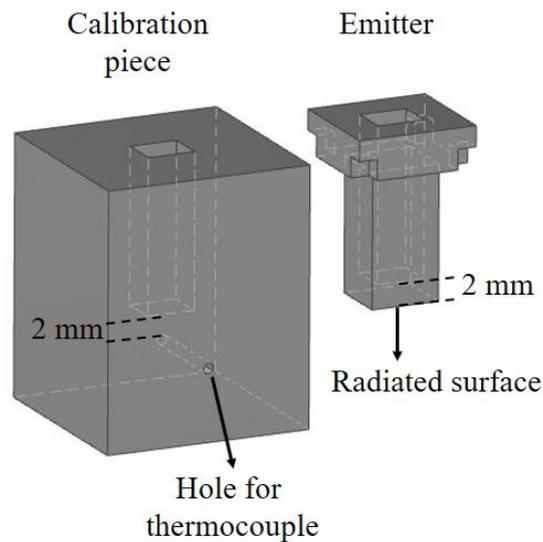

Figure 4: Sketch of the piece used for calibrating the temperature measurements (left) and the emitter (right).

The calibration piece was thermally isolated while heating up (by wireless power transmission using the RFLA500D laser) in order to homogenize its temperature. Under steady-state conditions, the emissivity parameter of the pyrometer was adjusted until the pyrometer and the thermocouple provided the same temperature measurement (±1 ºC). This was carried out at temperatures in between 450 ºC (lower limit of the pyrometer measurement) and 1000 ºC (upper limit of the thermocouple measurement). Above 855 ºC a constant value for the emissivity parameter (0.58) provided calibrated measurements, so this parameter was setup in the pyrometer for future measurements.

Although the emitter temperature is measured 2 mm above the surface that illuminates the TPV cell, regarding the thermal conductivity of the emitter [13] and the net heat fluxes stablished in this work, the difference expected between the temperature of the surface measured and the temperature of the radiated surface is lower than the temperature uncertainties presented in Table I.

**ACKOWLEDGEMT**

The authors gratefully thank Antora Energy for providing the InGaAs cells tested in this work. The authors also acknowledge Dr. Pablo García-Linares for the fruitful discussions. This work has been funded by the project GETPV (PID2020-115719RB-C22) funded by MCIN/AEI/10.13039/501100011033; by the project MADRID-PV2 (S2018/EMT-4308) funded by the Comunidad de Madrid with the support from FEDER Funds; and by the project NANOSPACE (APOYO-JOVENES-21-E641SS-135-LEV393) funded by the Comunidad de Madrid through the call Research Grants for Young Investigators from Universidad Politécnica de Madrid.